\documentclass [11pt]{amsart}

\usepackage{graphicx}
\usepackage{amssymb}
\usepackage{amsmath}
\usepackage{enumerate}

\newtheorem{definition}{{\bf Definition}}[section]
\newtheorem{theorem}[definition]{{\bf Theorem}}

\newtheorem{remarks}[definition]{\noindent {\bf Remarks}}

\newtheorem{lemma}[definition]{\noindent {\bf Lemma}}
\def\endproof{\hfill {\kern 6pt\penalty 500
\raise -0pt\hbox{\vrule \vbox to5pt {\hrule width 5pt
\vfill\hrule}\vrule}}}
\def\centerpicture #1 by #2 (#3){\leavevmode
        \vbox to #2{
        \hrule width #1 height 0pt depth 0pt
        \vfill
        \special{pictfile #3}}}
\begin{document}
\title{N-free extensions of posets.   Note on a theorem of P.A.Grillet.}
\author{Maurice Pouzet*}
\thanks{*Supported by CMCU}
\address{ PCS, Universit\'e Claude-Bernard Lyon1,
 Domaine de Gerland -b\^at. Recherche [B], 50 avenue Tony-Garnier, F$69365$ Lyon cedex 07, France}
\email{pouzet@univ-lyon1.fr }
\author{Nejib  Zaguia }

\address {SITE,  Universit\'e d'Ottawa, 800 King Edward Ave, Ottawa, Ontario, K1N6N5,  Canada}
\email{zaguia@site.uottawa.ca}

\maketitle
\date{\today}
\begin{abstract} Let $S_{N}(P)$ be the poset obtained by adding a dummy  vertex on each diagonal edge of the $N$'s of a finite poset $P$. We show that $S_{N}(S_{N}(P))$ is $N$-free.  It follows that  this poset  is  the smallest $N$-free barycentric subdivision of the diagram of $P$, poset  whose existence  was proved by P.A. Grillet.  This is also the poset obtained by the algorithm starting with $P_0:=P$ and  consisting at step $m$ of adding  a dummy vertex on a  diagonal edge of some $N$ in $P_m$, proving that the result of this  algorithm does not depend upon the particular choice of the diagonal edge choosen at each step.  These results are linked to  drawing of posets.\end{abstract}



\section{Introduction}

 An $N$ is a  poset made of four vertices  labeled $a,b,c,d$ such that   $a<c,b<c, b<d$,   $b$ incomparable to $a$, $a$ incomparable to $d$ and $d$ incomparable to $c$ (see Figure 1(a)). This simple poset plays an important role  in the algorithmic  of posets \cite {rival}. It  can be contained in a poset $P$ in essentially two ways, leading to the characterization of two basic  types of posets, the {\it series-parallel} posets and the {\it chain-antichain complete} (or C.A.C) posets. 
 
  The first way is related to the comparability graph of $P$.  An $N$ can be contained in $P$ as an induced  poset, that is $P$ contains four vertices on which the comparabilities are those indicated above.  Finite posets  with no induced $N$ are called {\it series-parallel}. Indeed, since their comparability graph contains no induced $P_4$ (a four vertices path ) they can obtained from the one element poset by direct and complete sums (a  result which goes back to Sumner \cite{sumner}, see also \cite{valdes}). The second way is  related to the (oriented) diagram of $P$. This is the object of this note.
   
In order to describe this other way, let us recall that a   {\it covering pair} in  a poset $P$ is a pair $(x,y)$ such that  $x<y$ and there is no $z\in P$ such that $x<z<y$. The (directed){\it diagram } of $P$ is the directed graph,  denoted  by  $Diag(P)$, whose vertex set is $P$ and edges are the covering pairs of $P$ . If $(x,y)$ is a covering pair, 
we  say that $x$ {\it is covered by }$y$, or $y$ {\it covers } $x$, a fact that we denote $x\prec_P y$, or $x\prec y$ if there is no risk of confusion, or $(x,y)\in Diag(P)$. We denote by $Inc(P)$ the set of pairs $(x,y)$ formed of incomparable elements. 
\begin{definition}\label{def}
Let   $a,b,c,$ and $d$ four elements of $P$, we say that these elements form:
\begin{enumerate} 
\item an $N$ in $P$ if $b\prec c$, $a\prec c$, $b\prec d$, and $(a, d)\in Inc(P)$;
\item an $N'$ in $P$ if $b\prec c$, $a< c$, $b<d$, and $(a, d)\in Inc(P)$;
\item an $N$ in $Diag(P)$ if $b\prec c$, $a\prec c$, $b\prec d$, and $a\not \prec d$;
\end{enumerate}
\end{definition}An $N$ in $P$ is also an  $N'$ in $P'$ and, provided that $P$ is finite, an $N'$ in $P$ yields an $N$ $\{a',b,c,d'\}$in $P$.
An $N$ in $P$ induces an $N$ in $Diag(P)$; the converse is false: if $\{a,b,c, d\}$ is an $N$ in $Diag(P)$ as above,  then $a< d$ is possible, but  -provided that $P$ is finite-    the $4$ element subset $a', a, c,b$, where $a\prec a'\leq d$, is an $N$ in $P$. Thus,  if $P$ is finite, it contains an $N$ under one of these three forms if it contains all.  We say that $P$ is $N$-{\it free} if it contains no $N$. 
It was proved by P.A.Grillet \cite {grillet} that {\it a finite poset $P$ is $N$-free if and only if $P$ is chain-antichain complete (or C.A.C) that is if every maximal chain of $P$ meets every maximal antichain of $P$ }(the formulation  $N$-free in terms of the $N$ defined in $1)$  is due to Leclerc and Monjardet \cite{leclerc}). 
 \begin{figure}[htbp]
  \centering 
\includegraphics[width=3.2in]{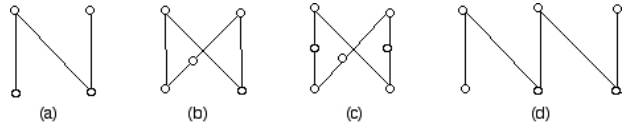}
  \caption{Examples of posets containing an $N$.}
  \label{fig3-nejib}
\end{figure}
   \begin{figure}[htbp]
  \centering 
 \includegraphics[width=3.2in]{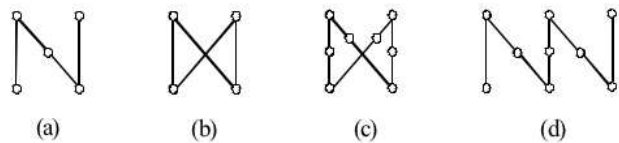}
  \caption{Examples of $N$-free posets}
  \label{fig2-nejib}
\end{figure}

A {\it barycentric subdivision} of the diagram of a poset $P$ consists to add finitely many vertices, possibly none, on each edge of the diagram of $P$. These vertices added to those of $P$ provides a new poset in which $P$ is embedded. We denote by $S(P)$ the poset obtained by adding just one vertex on each  edge of the diagram of $P$.  As it is immediate to see, this poset is $N$-free. In his embedding theorem (Theorem 7 \cite{grillet}) P.A.Grillet proves that among the $N$-free posets obtained as barycentric subdivisions of a finite poset $P$ there is one, denoted $\overline P$, which is minimal.  In this note, we provide a simple description of $\overline P$ and  give some  consequences. 

If $A:= \{a, b, c, d\}$ is an $N$ in $P$ (resp. a $N$ in $Diag(P)$), as in Definition \ref{def}, we say that   the pair $(b,c)$ is the {\it diagonal edge} of this  $N$. Let  $N_{diag}(P)$ be  the set of diagonal edges of all the  $N$'s in $P$ and let $S_{N}(P)$ be the poset obtained by adding a dummy  vertex on each  edge in $N_{diag}(P)$. 
\begin{theorem}  \label{maintheorem1}Let $P$ be a finite poset. Then $S_{N}(S_{N}(P))$ is $N$-free. In fact this is the smallest $N$-free poset $\overline P$  which  comes from a barycentric subdivision of $Diag(P)$. 
\end{theorem}
 This result translates  to an algorithm which transforms a poset into an 
 $N$-free poset: execute twice the algorithm consisting to add simultaneously a vertex on each $N$ of a poset.  Figure \ref{fig3-nejib} shows an execution of this algorithm. Two dummy elements 6 and 7 are created during the first execution. Another two, 8 and 9, are produced during the second execution. After the second execution, the resulting poset  does not contain an $N$. 
  \begin{figure}[htbp]
  \centering 
 \includegraphics[width=4in]{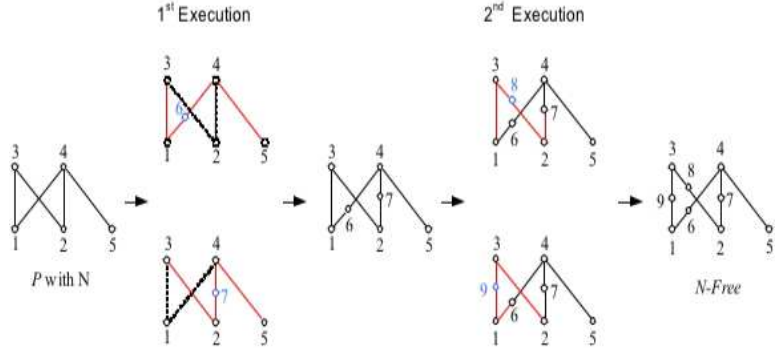}
  \caption{Execution of the  algorithm}
  \label{fig3-nejib}
\end{figure}
Instead of adding simultaneously the dummy vertices, we may add them successively.
\begin{theorem} \label{maintheorem2}The algorithm starting with $P_0:=P $ and adding at step $m$ a dummy vertex on a diagonal edge of some  $N$ in $P_m$ stops on $\overline P$. Hence the result and the number of steps does not depends upon the particular choice of the diagonal edges choosen at each step.
\end{theorem}
 \begin{remarks}
\begin{enumerate}
\item If instead of the diagonal edges of $P$ we consider those of $Diag(P)$, one  get the same conclusion as in Theorem  \ref{maintheorem1} and Theorem \ref {maintheorem2}  (see Remark \ref{rmk}  below ).
\item  A poset $P$  can be embedded into an $N$-free poset which does not come from a barycentric extension of its diagram, but a minimal one is not necessarily isomorphic to $\overline P$ or to a quotient of $\overline P$.   The posets represented in $(b)$ and $(d)$ of Figure 4 are minimal $N$-free extensions of $ A$ and $B$;  the first one is a quotient of $\overline A$ represented in $(a)$ the second is not a quotient of $\overline B$ represented in $(c)$. 
\item P.A. Grillet considered infinite posets satisfying some regularity condition.  We restricted ourselves to finite posets. How our results translate to the infinite? 
\end{enumerate} 
\end{remarks}  \begin{figure}[htbp]
  \centering 
 \includegraphics[width=3in]{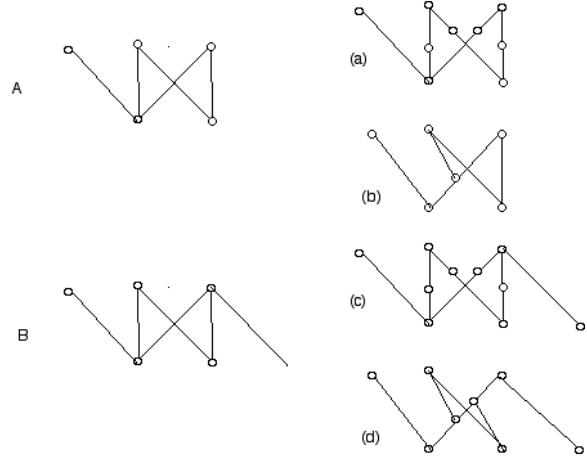}
  \caption{Minimal $N$-free extensions}   \label{fig4-nejib}
\end{figure}
The motivation for this research came from drawing of posets. A good drawing solution that works for all posets  is clearly out of reach. However, if every  poset can be embedded into another with a particular structure, and at the same time these particular structures can be nicely drawn,  then this can lead to an interesting approximation of general ordered set drawing.
In \cite{rakoto} was presented an approach for drawing $N$ -free posets. The  algorithm, called {\it LR-drawing} (LR for left-right),  consists of three steps: The first step is to convert $P$  into an $N$-free poset $Q$. The second step is to apply the LR-drawing to $Q$.  The third and last step is to retrieve $P$ from the drawing of $Q$. The first part of the algorithm requiring to look at the possible extensions of a poset into an $N$-free one, this suggested an other look at the barycentric extensions of a poset and lead to the present results.

\section{Proofs}
In this section, we consider a {\it finite poset} $P$. A basic ingredient of the proofs is the 
set  $A(P)$  of  $(b,c)\in Diag(P)\setminus N_{diag}(P)$ such that  there are two vertices $a,d\in P$ such that $a< c, b<d$, $(a,b), (c,d)\in Inc(P)$,  and either $(a, c)\in N_{diag}(P)$ or $(b,d)\in N_{diag}(P)$. 
In our definition of members of $A(P)$, we could  have supposed $a\prec c$ and $b\prec d$. The definition we choose is closer to the one  considered in Lemma 11 of Grillet's paper. 

An important feature  of a barycentric subdivision is that each new element  has a unique upper cover and a unique lower cover. This fact is at the root of the following lemma.

\begin{lemma}\label {newedge} Let $P'$  be a barycentric subdivision of $P$ and $a,b,c,d\in P'$. If $a<c, b<d, (b,c)\in Diag(P')$ and  $(a,b), (d,c)\in Inc (P')$ then $b,c\in P$; if, moreover, $(a, c), (b,d)\in Diag(P')$ and $a<d$ then $a,d\in P$. \end{lemma}

\begin{proof} If $b$ or $c$ is not in $P$ then $(b,c)$ is  a new edge, hence  either $b$, or $c$, is a dummy vertex. If $b$ is a dummy vertex, we have $c<d$, whereas if $c$ is a dummy vertex we have  $a<b$,  contradicting our hypothesis. If $a\not\in P$ then $(a,c)$ is a new edge and thus $a$ is a dummy vertex on some edge $(a', c)\in Diag(P)$;   from $a<d$,  we get $c<d$, a contradiction. Applying this to the dual poset $P^{dual}$ we get $d\in P$. 
\end{proof}

\begin{lemma} \label{first}Let $\{a, b,c, d\}$ four elements of $P$ such that $(a, c), (b,d)\in Diag(P)$, $(b, c)\in Diag(P)\setminus N_{diag}(P)$. 
\begin{enumerate}
\item  $a<d$ and if $(a,d)\not \in Diag(P)$ then $(a,c), (b,d)\in N_{diag}(P)$; 
\item If  $(a,c)\in  N_{diag}(P)$ then 
\begin{enumerate}
\item $(x, b)\in Inc(P)$  for every $x\in P$ such that $(a, x)\in $Diag (P)$ $  and $\{a, c, x, y\}$ 
 witnesses  the fact that $(a,c)\in  N_{diag}(P)$ for some $y\in P$;
 \item $(a,d)\in N_{diag} (P)$ iff $(a,d)\in Diag(P)$. 
 \end{enumerate}
\item  $(a,c)\in  N_{diag}(P)$  if and only if there is some $x\in P$ such that $(a,x)\in Diag(P)$ and $(x, b)\in Inc(P)$.
\end{enumerate} 
\end{lemma} 
\begin{proof}
$(1)$ If $a\not < d$ then $\{a, b, c, d\}$ is an $N$ in $P$ hence $(b,c)\in N_{diag}(P)$ contradicting the fact that $(b, c)\in Diag(P)\setminus N_{diag}(P)$. Let $x\in P$ such that $a\prec x\leq d$. Then $\{x, a, c, b\}$ is an $N$ in $P$ hence $(a,c)\in N_{diag}(P)$. With this argument applied to $P^{dual}$ we get $(b,d)\in N_{diag}(P)$. 

$(2)$ Suppose $(a,c)\in N_{diag}(P)$. $(i)$. Let $x, y$ such that   $(a,x), (y, c)\in Diag(P)$ such that $\{x, a, c, y\}$ witnesses that $(a, c)\in N_{diag}(P)$. If $(x, b)\not \in Inc(P)$ then $b<x$. Let $b'\in P$  such that $b\prec b'\leq x$. Then  $\{y, c, b', d\}$ is an $N$ in $P$ thus $(b, c)\in N_{diag}(P)$ contradicting our hypothesis. $(ii)$. Suppose $(a,d)\in Diag(P)$. Let $x, y$ as above. Since $(x, b)\in Inc(P)$, $\{x,a,d, b\}$ is an $N$ in $P$, hence $(a,d)\in N_{diag}(P)$. The converse is obvious. 

$(3)$ follows immediately from $(2-a)$. 
\end{proof}

\begin{lemma}\label{A(P)}$N_{diag}(S_N(P))= A(P)$
\end{lemma}
\begin{proof} Set $P':= S_N(P)$.
 
$(a)$ $N_{diag}(P')\subseteq A(P)$.  Let $(b,c)\in N_{diag}(P')$.

{\bf Claim 1}  $(b, c)\in Diag(P)\setminus N_{diag}(P)$. Moreover, if $A:= \{a, b, c, d\}$ is an $N$ in $P'$  with $a\prec_{P'}c$ and $b\prec_{P'}d$ then   $a$  or $d$ are in $P' \setminus P$.

{\bf Proof of  Claim 1} According to Lemma \ref{newedge}  we have $b,c\in P$. Since $(b,c)\in Diag(P')$,  it follows  $(b, c) \in Diag(P)\setminus N_{diag}(P)$. Since $b,c\in P$, if  $a$ and $d$  are in $P$ then   $\{a,b,c,d\}$ is an $N$ in $P$ and thus $(b, c)$ has been subdivided, hence $(b,c)\not\in Diag(P')$ a contradiction. \endproof

Let $A$ as above. 

{\bf Case 1}.  $a\in P'\setminus P$.  In this case $a$ is a dummy vertex on some edge $(a', c)\in N_{diag}(P)$.  Since $(b, d)\in Diag (P')$ there is some $d'\in P$ such that $b\prec_Pd'$ and $d\leq d'$ ($d'=d$ if $d\in P$, otherwise $(b, d)\in Diag(P')$ in which case $d$ is a dummy vertex on $(b, d')$). Thus $A':= \{a', b, c, d'\}$ witnesses the fact that $(b,c)\in A(P)$.

{\bf Case 2}. $d\in P'\setminus P$. 
This case reduces to Case $(1)$ above by considering the dual poset $P^{dual}$.  
From Claim 1 there is no other case. The proof of $(a)$ is complete. 

$(b)$ $A(P)\subseteq N_{diag}(P')$. Let $(b,c)\in A(P)$. Let $\{a,b,c,d\}$,  with $(a,c), (b,d)\in Diag (P)$,   witnessing it. If $(a, c)\in N_{diag}(P)$, let $u$ be a dummy vertex on $(a, c)$ then $\{u, c, b, d'\}$, where  $d':= d $ if $(b, d)\not \in N_{diag}(P)$ and $d'$ is a dummy vertex on $(b,d)$ otherwise, is an $N$ in $P'$ hence $(b,c)\in N_{diag}(P')$. If $(b,d)\in N_{diag}(P)$,  apply the above case to $P^{dual}$.  \end{proof}

\begin{lemma} \label{smallemma} $A(S_N(P))=\emptyset$ \end{lemma}
\begin{proof}
Suppose the contrary. Set  $P':= S_N(P)$ and let $(b,c)\in A(P')$.  Let $A:= \{a,b,c,d\}$,  with $(a,c), (b,d)\in Diag (P')$, witnessing the fact that $(b,c)\in A(P')$.  According to $(1)$ of Lemma \ref{first} applied  to $P'$, we have $a<d$. Thus from Lemma  \ref{newedge}, we have $a,b,c,d\in P$. 

{\bf Case 1}. $(a,c)\in N_{diag}(P')$. According to $(3)$ of Lemma \ref{first} applied to $P'$ there is some $x\in P'$ such that $(a,x)\in Diag(P')$ and $(x, b)\in Inc(P')$. 

Next,  $x\in P'\setminus P$. Indeed,  $\{x,a,c,b\}$ is an $N$ in $P'$. Thus,  if $x\in P$,  this is  an $N$ in $P$  and $(a,c)\in N_{diag}(P)$, hence a dummy vertex is added on $(a,c)$ in $P'$ contradicting $(a,c)\in Diag(P')$. Finally, we consider two subcases:

{\bf Subcase 1.1}. $(a,d)\in Diag(P')$. In this case,  $(x, d)\in Inc(P)$ and, since $x\not \in P$, $(a,x)\in Diag(P')\setminus Diag(P)$. Hence, there is $x'\in P$ such that $x$ is a dummy vertex of $(a,x')\in N_{diag}(P)$.  Let $A':= \{x', a, c, b\}$. We have $(a,c), (b,c)\in Diag(P)\setminus N_{diag}(P)$ and $(a,x')\in N_{diag}(P)$. Thus $(a,c)\in A(P)$. According to $(1)$ of Lemma \ref{first} $(b,x')\in Diag(P)$. Next, according to $(3)$ of Lemma \ref{first}, there is some $v\in P$ such that $(v,x')\in Diag(P)$ and $(v,c)\in Inc(P)$. 
It follows that $\{v,x', b, c\}$ is an $N$ in $P$ hence $(b,x')\in N_{diag}(P)$.  If $b'$ is  a dummy vertex on $(b,x')$ then $\{b', b,c, a\}$ is an $N$ in $P'$ hence $(b,c)\in N_{diag}(P')$ contradicting $(b,c)\in A(P')$. Thus this subcase leads to a contradiction. 

{\bf Subcase 2.2}. $(a, d)\not \in Diag (P')$. 
In this case, we may suppose  $x<d$. In fact 
$(x,d)\in Diag(P')$. Indeed, if $(x,d)\not \in Diag (P')$ then there is $d'\in P$ such that $x<_{P'}d'\prec_{P}d$. But, then $\{d',d,  c, b\}$ is an $N$ in $P$, thus $(a,c)\in N_{diag}(P)$ proving that $(a,c)\not \in Diag (P')$ a contradiction. It follows that $x$ is a dummy vertex added on $(a,d)\in N_{diag}(P)$. Since 
$(a,d)\in N_{diag}(P)$, $(b,d)\in Diag(P)\setminus N_{diag}(P)$ and $(b,c)\in Diag(P)$, $(b,d)\in A(P)$. Since $(a,c)\in Diag(P)$ it follows from $(2-b)$ of Lemma \ref{first}  that $(a,c)\in N_{diag}(P)$ contradicting $(a,c)\in Diag (P')$. This subcase leads to a contradiction too.

{\bf Case 2}.  $(b,d)\in N_{diag}(P')$. This case reduces to the previous one by considering the dual poset $P^{dual}$. Hence, it leads to a contradiction. 

Consequently $A(P')=\emptyset$. The proof is complete. 
\end{proof}

{\bf Proof of Theorem  \ref{maintheorem1}}.
Let $P':= S_N(S_N(P))$. According to Lemma \ref{A(P)} and Lemma \ref{smallemma},  $N_{diag}(P')=A(S_N(P))=\emptyset$. Clearly, an $N$-free poset $Q$ associated with a barycentric subdivision of $Diag(P)$ must include $S_N(P)$. This applied to $S_N(P)$ gives that $Q$ contains $P'$. Hence $P'$ is the smallest $N$-free poset coming from a barycentric subdivision of $Diag (P)$, thus it coincides with the poset $\overline P$ constructed by P.A.Grillet. \endproof

\begin{lemma} \label{second}Let $P'$ with $P\subseteq P'\subseteq S_N(S_N(P))$; then $N_{diag}(P')\subseteq N_{diag}(P)\cup A (P)$. 
\end{lemma}
\begin{proof}
Let $(b,c)\in N_{diag}(P')$. Suppose  $(b,c)\not \in N_{diag}(P)\cup  A_(P)$. Let $Q:= S_N(S_N(P))$. We claim  that $(b,c)\in N_{diag}(Q)$. Let $A:= \{a,b,c,d\}$ be an $N$ of $P'$ witnessing the fact that $(b,c)\in N_{diag}(P')$. Since,  from Lemma \ref{A(P)} $(b,c)\not \in N_{diag}(P)\cup N_{diag}(S_N(P))$,  $(b, c)\in Diag(Q)$ thus $A':= \{ a', b, c, d'\}$ where $a\leq a'\prec_Q c$ and $b\prec_Q d'$ is an $N$ in $Q$ proving our claim.  Next, with  $(b,c)\in N_{diag}(Q)$ and $Q:= S_N(S_N(P))$,  we get from Lemma \ref{A(P)} that $(b,c)\in A( S_N(P))$. Since, from Lemma \label{smallemma}, $A(S_N(P))=\emptyset$, we get a contradiction. This proves the lemma.
\end{proof}

 {\bf Proof of Theorem  \ref{maintheorem2}} An immediate induction using Lemma \ref{second} shows that each $P_m$ is a subset of $Q:= S_N(S_N(P))$. Since $Q$ is the least $N$-free subset of $S(P)$ containing $P$ the algorithm stops on $Q$. The number of steps is the size of $N_{diag}(P)\cup A (P)$.\endproof 
 \begin{remarks}\label{rmk}
 \begin{enumerate}
 \item Let $ND_{diag}(P)$ be the set of  diagonal edges of the $N$'s  in $D(P)$.  Clearly,  $ND(P)\subseteq N_{diag}(P)\cup A (P)$. Thus,  with the same proof as for Theorem  \ref{maintheorem2}, we obtain that the algorithm consisting to add at  step $m$ a  dummy vertex on  an edge of some $N$ in $D(P_m)$ ends on $\overline P$. Similarly, with Lemma \ref{second} we get that $ND_{diag}(ND_{diag}(P))=\overline P$;
 \item The fact that the algorithm given in Theorem \ref{maintheorem2} stops is obvious: at each step, $P_m$ is a subset of $S(P)$. The fact that the number of steps in independent of the choosen edges is more significant. This suggests a deepest investigation. We just note that if $P_m$ contains just one $N$ then $P_{m+1}$ is $N$-free (we leave the proof to the reader). 
 
 \end{enumerate}
 \end{remarks}

\end{document}